\documentclass[10pt]{article}

\usepackage{amsmath}
\usepackage{amssymb}

\usepackage{graphicx}

\usepackage{cite}

\usepackage{color} 

\usepackage{tikz}
\usetikzlibrary{matrix}

\usepackage[labelfont=bf,labelsep=period,justification=raggedright]{caption}

\makeatletter
\renewcommand{\@biblabel}[1]{\quad#1.}
\makeatother

\date{}

\begin{document}

\begin{flushleft}
{\Large
\textbf{An analysis of the interaction between influenza and respiratory 
syncytial virus based on acute respiratory infection records}
}\\

Yendry N. Arguedas-Flatts$^{1}$, 
Marcos A. Capistr\'an$^{1,\ast}$, 
J. Andres Christen$^{2}$,
Daniel E Noyola$^{3}$
\\
\bf{1} Matem\'aticas B\'asicas, Centro de Investigaci\'on en Matem\'aticas A.C., Guanajuato, Gto, M\'exico
\\
\bf{2} Probabilidad y Estad\'istica, Centro de Investigaci\'on en Matem\'aticas A.C., Guanajuato, Gto, M\'exico
\\
\bf{3} Departamento de Microbiolog\'ia, Facultad de Medicina, Universidad Aut\'onoma de San Luis Potos\'i, San Luis Potos\'i, SLP, M\'exico
\\

$\ast$ E-mail: Corresponding marcos@cimat.mx
\end{flushleft}

\section*{Abstract}
Under the hypothesis that both influenza and respiratory syncytial virus (RSV) are 
the two leading causes of acute respiratory infections (ARI), in this paper we have used a standard two-pathogen epidemic model as a regressor to explain, on a yearly basis, high season ARI data in terms of the contact rates and initial conditions of the mathematical model. 
The rationale is that ARI high season is a transient regime of a noisy system, 
e.g., the system is driven away from equilibrium every year by fluctuations 
in variables such as humidity, temperature, viral mutations and human behavior.
Using the value of the replacement number as a phenotypic trait associated to fitness, we provide evidence that influenza and RSV coexists throughout the ARI high season through superinfection.


\section*{Introduction}
It is known that influenza and respiratory syncytial virus (RSV) are 
the two leading causes of acute respiratory infections 
(ARI)~\cite{grondahl1999rapid,fleming2005respiratory}, 
their interaction however is complex and numerous research efforts have
been conducted towards elucidating the underlying ecological and epidemic mechanisms
of ARI~\cite{dietz1979epidemiologic,huang2005dynamical}, and the causes of
seasonality in ARI~\cite{denny1987acute,eccles2002explanation}. It has long been argued that there is viral interference between influenza and respiratory syncytial 
virus~\cite{anestad1982interference,rohani1998population,linde2009does,anestad2009interference}. 
More broadly, interference of viral replication among viruses has been suggested for many years by epidemiological, clinical and laboratory studies~\cite{anestad1982interference,aanestad1987surveillance,wagner1963cellular,
dianzani1975viral,cowling2012increased,seppala2011viral,ge2012evaluating}. Reduced susceptibility to a second viral infection during convalescence has been described. In addition to a reduction in exposure as a result of a decrease in social contacts while a person is sick and during 
convalescence~\cite{rohani2003ecological}, mediators with antiviral activity (such as interferon) are produced by cells of the infected host; these render the individual less susceptible to secondary viral infections. In vitro studies have shown that interferon production reduces the ability of viruses to infect cells in 
cultures~\cite{wagner1963cellular}. Epidemiological studies have shown that peak influenza and RSV activity tends to occur at different time 
periods~\cite{aanestad1987surveillance,nishimura2005clinical}. On occasions 
widespread circulation of a virus may prevent the emergence of another virus. For instance, during the 1975-76 and 1979-80 winter seasons RSV circulation was low in Norway; it was suggested that this was the result of interference by influenza activity in the region~\cite{aanestad1987surveillance}. Also, a change in RSV epidemiology was observed during 2009 in France concurrent with circulation of the influenza A(H1N1) 2009 virus~\cite{casalegno2010impact}.
Also, the theoretical study of 
the ecological phenomena that viral interference may give rise to (competitive exclusion, coexistence, coevolution, etc) have received considerable attention~\cite{levin1981selection,bremermann1989competitive,castillo1998relationship,
feng1997competitive,hochberg1990coexistence,may1994superinfection}. 
On the other hand, among the possible causes of seasonality of respiratory infectious diseases are the following: 
during adverse climatic conditions (colder season in temperate climates; rainy season in tropical areas) people spend more time indoors, and increased aggregation leads to closer contact between susceptible and infected individuals
~\cite{weber1998respiratory}; colder 
temperatures correspond to drier air, which dehydrates mucose membranes
making people more vulnerable to infection~\cite{lowen2007influenza,shaman2009absolute}. 
Another feature of seasonal epidemics of respiratory infections that has attracted
interest from researchers is the development of methods for early detection of 
epidemic outbreaks~\cite{unkel2012statistical}. Robust outbreak detection methods 
are mandatory in order to identify the onset of pandemic influenza. However,
non-pandemic outbreaks have striking regularity in large cities, and the onset of ARI 
high season corresponds to approximately the first time that ARI records hit its 
historical mean (sometime during the last quarter of a given year in the northern
hemisphere). Beyond the identification of epidemic outbreaks it is necessary, from 
the point of view of public health intervention, to discriminate between illnesses 
caused by influenza and those caused by RSV in order to define appropriate management 
routines~\cite{zambon2001contribution}.

Aiming at contributing towards methods to study the interaction of influenza and 
RSV, in this paper we have used a standard two-pathogen 
epidemic model as a regressor to explain, on a yearly basis, high season ARI data in 
terms of the contact rates and initial conditions of the mathematical model. 
The rationale is that ARI high season is a transient regime of a noisy system, 
e.g., the system is driven away from equilibrium every year by fluctuations in variables such as humidity, temperature, viral mutations and human behavior. 
The effect of ARI caused by other pathogens is not modeled explicitly. Using 
the replacement number as a phenotypic trait to define fitness, we provide evidence that influenza and RSV coexists throughout the ARI high season through superinfection.

\section*{Materials and Methods}

\subsection*{Ethics statement}
This study included retrospective analyses of information available from databases (weekly number of ARI consultations and weekly number of viral detections). Weekly data for ARI consultations is recorded as part of routine surveillance activities carried out by the State Public Health Department. Virological information was derived from research projects carried out to analyze the epidemiology of viral respiratory infections and as part of the hospital's infection control program; the research projects were approved by the Research and Ethics Committee at Hospital Central "Dr. Ignacio Morones Prieto" and written informed consent was obtained from children's parents.

\subsection*{Data} 

\begin{figure}[!ht]
\begin{center}
\includegraphics[scale=0.75]{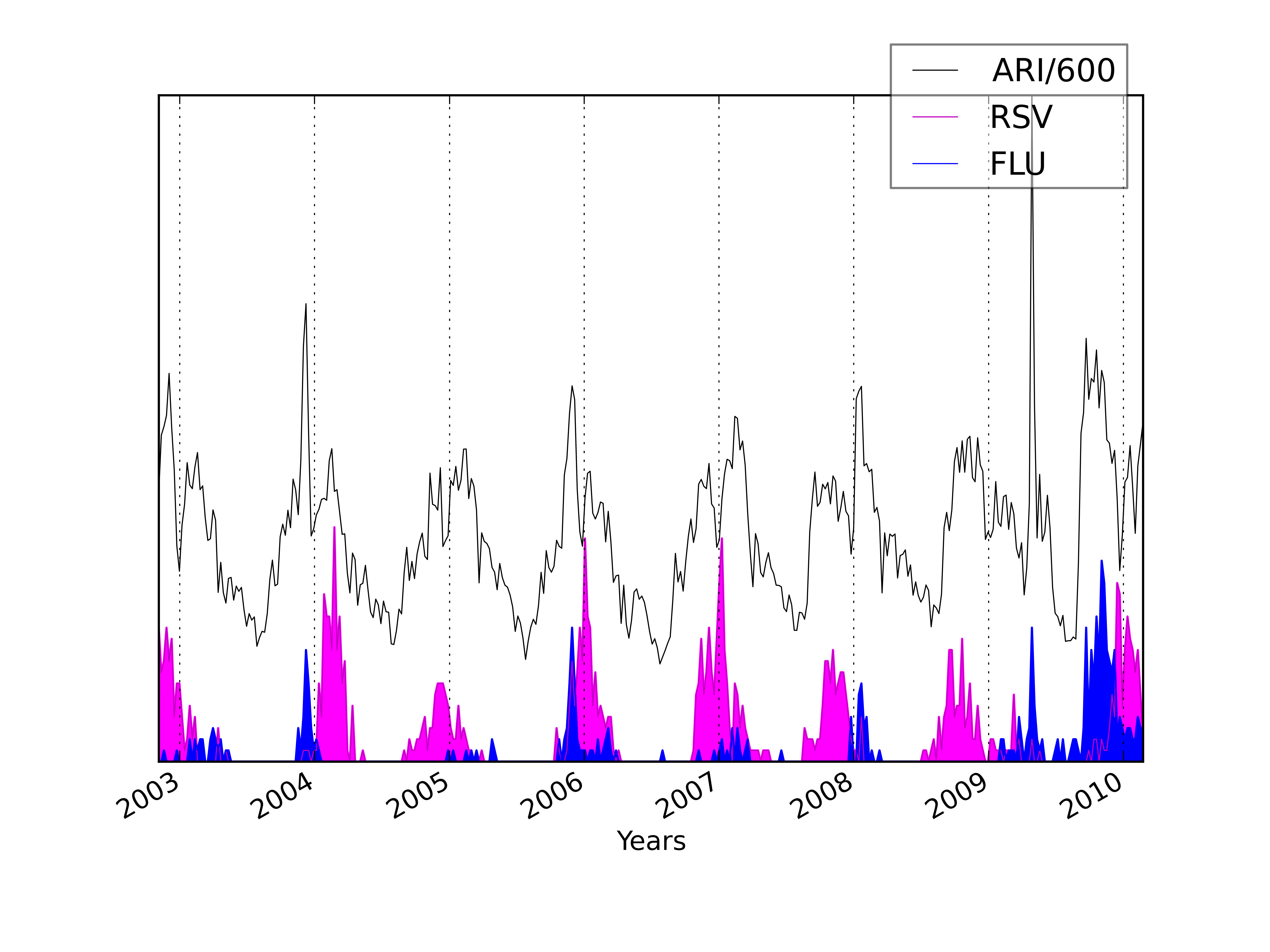}
\end{center}
\caption{{\bf Correlation in disease records.} Influenza and RSV are typically 
detected within a sentinel program at Hospital Central ``Dr. Ignacio Morones Prieto" 
in San Luis Potos\'i, M\'exico around the same time when ARI data from the state 
public health services reach its local maxima. Time series are shown in arbitrary scale.}
\label{fig:data}
\end{figure}

Data consists of weekly number of ARI consultations and weekly number of viral 
detections from the State of San Luis Potos\'i, Mexico ranging from 2002 to 2010. 
The seasonal trend of ARI and virus circulation are shown in Figure~\ref{fig:data}. 
Temperature and humidity data of San Luis Potos\'i was obtained from the National 
Climatic Data center http://www7.ncdc.noaa.gov/CDO/cdo.  

\subsection*{Epidemic outbreak}
There are methods to identify the onset of epidemics ranging across all subfields
of statistics~\cite{unkel2012statistical}. However, non-pandemic records of ARI
have a striking regularity in large cities, this is the case in the State of San Luis Potos\'i, which has an approximate population of 2,000,000 inhabitants. For the sake of simplicity, in this paper we examine ARI data corresponding to non-pandemic years.
Consequently, we regard the historical mean of ARI as the baseline, and define ARI 
high season as those weeks from October-November of a given year, when data reaches 
its historical mean, until April-May of the following year when data drops again 
below its historical mean, see supplementary material. 

\subsection*{Two-pathogen epidemic model}
\begin{figure}[!ht]
\begin{center}
\includegraphics[scale=0.75]{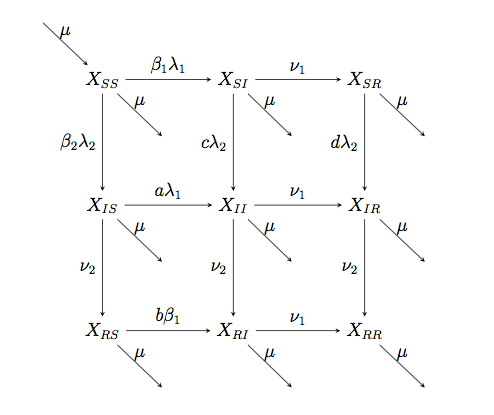}
\end{center}
\caption{{\bf Two-pathogen SIR model.} The quantities
$\lambda_{1}=(x_{SI}+x_{II}+x_{RI})/N$, 
$\lambda_{2}=(x_{IS}+x_{II}+x_{IR})/N$, $\nu_{1}$ and $\nu_{2}$ multiplied
by the corresponding state variable denote the propensity of each kind of event,
e.g. birth, infection with one disease, recovery or death.}
\label{fig:model}
\end{figure}

The two-pathogen SIR epidemic model considered in this paper
is defined by the graph shown in Figure~\ref{fig:model}. 

Model defined by Figure~\ref{fig:model} is a history-based 
model of two-pathogen epidemics, and no relation is made for the time being 
between the indexes 1,2 and the names of the diseases. Initially, we carry out a 
classical qualitative analysis of the deterministic equations~(\ref{eq:mean-field})
in order to distinguish between four noteworthy ecological regimes
(competitive exclusion, cocirculation, superinfection and coexistence of two 
pathogens in a population), which are further described below. 
Model~(\ref{eq:mean-field}) is the same model analyzed 
in~\cite{shrestha2011statistical}, except that we have neglected the convalescent 
state. Furthermore, state variables in model~(\ref{eq:mean-field}) 
account for all infections, from ARI to mild or asymptomatic infections. 
Consequently, in subsection {\bf Inference} we develop a statistical model
to compare the model prediction to ARI time series data under the hypothesis
that the number of ARI is proportional to the total number of infections caused 
by both diseases. Our aim is twofold,
on the one hand we want to identify an ecological regime implicit in 
model~(\ref{eq:mean-field}) that best explains
ARI data, on the other hand we want to estimate the contact rate of each 
disease in order to compute the replacement number of both diseases throughout
the ARI high season.

The deterministic equations for the dynamics of the state variables are

\begin{equation}
\label{eq:mean-field}
\begin{split}
\frac{dX_{SS}}{dt} &= \mu N - 
(\beta_{1}\lambda_{1} + \beta_{2}\lambda_{2}+\mu)X_{SS}\\
\frac{dX_{SI}}{dt} &= \beta_{1}\lambda_{1}X_{SS} 
- (\nu_{1}+\mu+c\lambda_{2})X_{SI}\\
\frac{dX_{SR}}{dt} &= \nu_{1}X_{SI} -
(\mu + d\lambda_{2})X_{SR}\\
\frac{dX_{IS}}{dt} &= \beta_{2}\lambda_{2}X_{SS} -
(b\lambda_{1} + \mu + \nu_{2})X_{IS}\\
\frac{dX_{II}}{dt} &= c\lambda_{1}X_{SI} + b\lambda_{2}X_{IS} -
(\nu_{1} + \nu_{2} + \mu)X_{II}\\
\frac{dX_{IR}}{dt} &= d\lambda_{2}X_{SR} + \nu_{1}X_{II} -
(\mu + \nu_{2})X_{IR}\\
\frac{dX_{RS}}{dt} &= \nu_{2}X_{IS} -
(\mu + b\lambda_{1})X_{RS}\\
\frac{dX_{RI}}{dt} &= b\lambda_{1}X_{RS} + \nu_{2}X_{II} -
(\nu_{1} + \mu)X_{RI}\\
\frac{dX_{RR}}{dt} &= \nu_{1}X_{RI} + \nu_{2}X_{IR} -
\mu X_{RR},\\
\end{split}
\end{equation}

where the state variable $X_{ij}$ denotes the number of individuals in state
$ij$ for $i,j\in\{S,I,R\}$. We write $\lambda_{1}=(X_{SI}+X_{II}+X_{RI})/N$ and 
$\lambda_{2}=(X_{IS}+X_{II}+X_{IR})/N$ for clarity. Contact rates are given by
$\beta_{1}$, $\beta_{2}$, $a$, $b$, $c$ and $d$, while $\nu_{1}$, $\nu_{2}$ are
recovery rates. We denote both birth and death rate by $\mu$. In the remainder 
of the paper we have taken $\nu_{1}=\nu_{2}=1/7\;\mbox{day}^{-1}$, and 
$\mu=1/75\;\mbox{year}^{-1}$ fixed. The total population is $N=1,000,000$.

\paragraph{Model Analysis} 

The basic reproductive number of model~(\ref{eq:mean-field}) 
is given by 
\begin{equation}
\label{eq:R0}
R_0= \max\{R_{0,1},\,\, R_{0,2}\} 
\end{equation}

where

\begin{equation*}
R_{0,1}=\frac{\beta_1 }{\nu_1+\mu },\,\, R_{0,2}=\frac{\beta_2 }{\nu_2+\mu } \label{r0}
\end{equation*}

The disease free equilibrium (DFE)
\begin{equation}
\label{eq:dfe}
(N,0,0,0,0,0,0,0,0)
\end{equation} 
is locally stable if $R_{0}<1$ and unstable if $R_{0}>1$.
There are two semi-endemic equilibria, the first one (EE1) exists if 
$R_{0,1}>1$ and $X_{IS}=X_{IR}=X_{II}=0$ and is given by
\begin{equation}
\label{eq:ee1}
\left(\dfrac{N}{R_{0,1}},\dfrac{\mu }{\beta_1}(R_{0,1}-1)N,\dfrac{\nu_1}{\beta_1} (R_{0,1}-1)N,0,0,0,0,0,0\right) 
\end{equation}
Since the model is symmetric, the second semi-endemic equilibrium (EE2)
exists if $R_{0,2}>1$ and $X_{SI}=X_{RI}=X_{II}=0$ and is given by
\begin{equation}
\label{eq:ee2}
x_2=\left(\dfrac{N}{R_{0,2}},0,0,\dfrac{\mu}{\beta_2}(R_{0,2}-1)N,0,0,\dfrac{\nu_2}{\beta_2}(R_{0,2}-1)N,0,0\right) 
\end{equation}
We remark that the region where both diseases coexist admits complex dynamics and its 
detailed analysis escapes the purpose of this paper. We plan to report a more complete qualitative analysis elsewhere.

\begin{itemize}
\item If $a=b=c=d=0$, cross immunity is perfect and equations~(\ref{eq:mean-field}) model pure competitive exclusion between two diseases. 
\item If $c=d=0$ and $a\neq0$, $b\neq0$, or  If $a=b=0$ and $c\neq0$, $d\neq0$,
equations~(\ref{eq:mean-field}) model superinfection, e.g., if If $c=d=0$ and 
$a\neq0$, $b\neq0$, the model predicts that individuals with the second infection can
acquire the first infection but not the other way around.
\item If $a=c=0$ and $b\neq0$, $d\neq0$ 
equations~(\ref{eq:mean-field}) model a regime where both infections circulate in the population but no individual has both infections at once.
\item If $a\neq0$, $b\neq0$, $c\neq0$, $d\neq0$ equations~(\ref{eq:mean-field}) model a general two-pathogen epidemic model.
\end{itemize}

The replacement number for both diseases is defined as follows

\begin{align}
R_{1}=\frac{\beta_{1}\,X_{SS} + a\,X_{IS} + b\,X_{RS}}{N(\nu_{1} + \mu)},\\ 
R_{2}=\frac{\beta_{2}\,X_{SS} + c\,X_{SI} + d\,X_{SR}}{N(\nu_{2} + \mu)}
\end{align}

Since the fitness of both diseases changes considerably throughout the ARI high 
season, the replacement number is an appropriate phenotypical trait to analyze
the relative fitness of the two diseases in the two-pathogen epidemic 
model~(\ref{eq:mean-field}).

\subsection*{Inference}
It is known that symptomatic infections requiring medical attention comprise 
only a fraction of all infections occurring during an epidemic period.
Data of infected individuals consists of counts.  
Consequently, we assume that the ARI weekly reports $z_{i}$,  
are independent realizations of a Poisson distribution 
$Z_{i}$ $i=1,...,n$ whose mean is proportional to the integral 
of the incidence of any new infections between two observation 
times $t_{i-1}$ and $t_{i}$ predicted by the deterministic model, 
i.e

\begin{equation}
\label{eq:data_distribution}
Z_{i} \sim \mbox{Poisson}(K\mathcal{I}_{i}(\theta))
\end{equation}

where 
$$\mathcal{I}_{i}(\theta)=
\int_{t_{i-1}}^{t_{i}}((\beta_{1}+a+b)\lambda_{1}
+(\beta_{2}+c+d)\lambda_{2})X_{SS}dt
$$

$\theta=(\beta_{1},\beta_{2},a,b,c,d,X_{IS}(0),X_{SI},(0),K)$ are the model 
parameters to be inferred and 
$z=(z_{1},...,z_{n})$ are the accumulated ARI measurements. 
The proportionality constant $K$ accounts for the fact that the 
number of ARI reports is proportional to the number of individuals 
infected with either influenza or RSV. We substract the historical 
mean to ARI data prior to carrying out the inference process.

\paragraph{Likelihood} Assuming that the observations are independent between 
them, the joint distribution of the observed ARI cases
is a good approximation to the conditional probability 
$\pi(a\mid\theta)$, which would simply be defined by the product of the individual probability density functions of the observations:
  
\begin{equation}
\label{eq:likelihood}
  \pi(z\mid\theta) = 
  \prod_{i=1}^{n}\frac{e^{-K\mathcal{I}_{i}(\theta)}
  (K\mathcal{I}_{i}(\theta))^{z_{i}} }{z_{i}!}
\end{equation}

\paragraph{Informative priors} Since all model parameters are non-negative we propose Gamma prior distributions for them. For the parameters that account for
contact rate we chose the instrumental parameters in the Gamma distribution
such that its expected value corresponds to previous reports of $R_{0}$
for both influenza and respiratory syncytial 
virus~\cite{weber2001modeling,chowell2008seasonal}, e.g.
$$
\beta_{1}\sim\mbox{Gamma}(a_{\beta_{1}},b_{\beta_{1}}).
$$ where $a_{\beta_{1}}$ and $b_{\beta_{1}}$ are instrumental
parameters such that $E[\beta_{1}/(\nu_{1}+\mu)]=1.5.$
Likewise, for the initial conditions we have set
$X_{SI}(0)\sim \mbox{Gamma}(a_{X_{SI}},b_{X_{SI}})$ and 
$X_{IS}(0)\sim \mbox{Gamma}(a_{X_{IS}},b_{X_{IS}})$ with expected value 0.1
in both cases. For the sake of keeping a parsimonious model we have neglected the initial number of individuals in the compartments $X_{II}$, $X_{IR}$ and $X_{RI}$.  
This gives rise to an a posteriori distribution

\begin{equation}
\label{eq:posterior}
\pi(\theta\mid z)\propto
\prod_{i=1}^{n}\mbox{Poisson}(K\mathcal{I}_{i}(\theta))\times \mbox{Gamma}(a_{\beta_{1}},b_{\beta_{1}})\times...\times\mbox{Gamma}(a_{X_{IS}},b_{X_{IS}}) 
\end{equation}

\paragraph{Markov Chain Monte Carlo}
We have explored the posterior distribution $\pi(\theta\mid z)$ with the 
software for Markov Chain Monte Carlo t-walk by Christen and Fox~\cite{christen2010general}.

\section*{Results}

For the analysis and inference carried out in this paper we have used a simplified version of the two-pathogen model analyzed in~\cite{shrestha2011statistical} where 
it was shown, with likelihood-based inference, that the interaction of 
multi-patoghen systems can be inferred from noisy time series if there is enough 
information about the epidemiological and demographic underlying processes.

\begin{figure}[!ht]
\begin{center}
\includegraphics[scale=0.75]{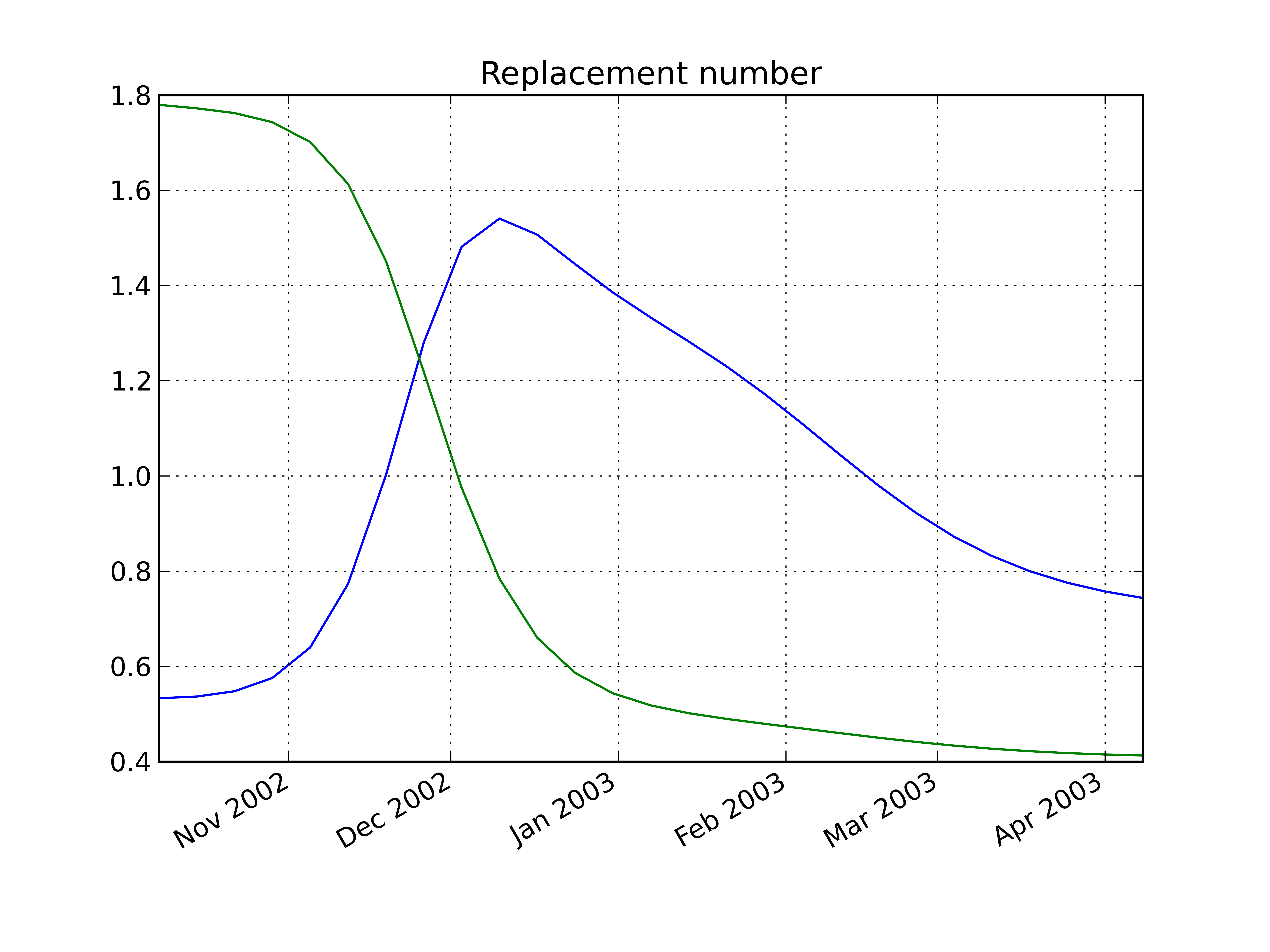}
\caption{{\bf Replacement number as a measure of fitness.} RSV (green line) has 
higher fitness at the 
beginning of the ARI high season. Although influenza initially has replacement 
number below 1, it remains present in the population and proliferates once the 
RSV epidemic outbreak starts to decline (Color online).}
\label{fig:replacement_number_2002_2003}
\end{center}
\end{figure}

The replacement number, see Figure~\ref{fig:replacement_number_2002_2003}, 
provides a line of evidence of our findings: intially RSV has higher replacement
number (fitness) and invades first. However, RSV fitness is monotonically 
decreasing while influenza starts with lower but monotonically increasing 
replacement number. Eventually influenza is the pathogen with the larger
replacement number, until it also reaches a peak and starts to decline. 

\begin{figure}[!ht]
\begin{center}
\includegraphics[width=\textwidth]{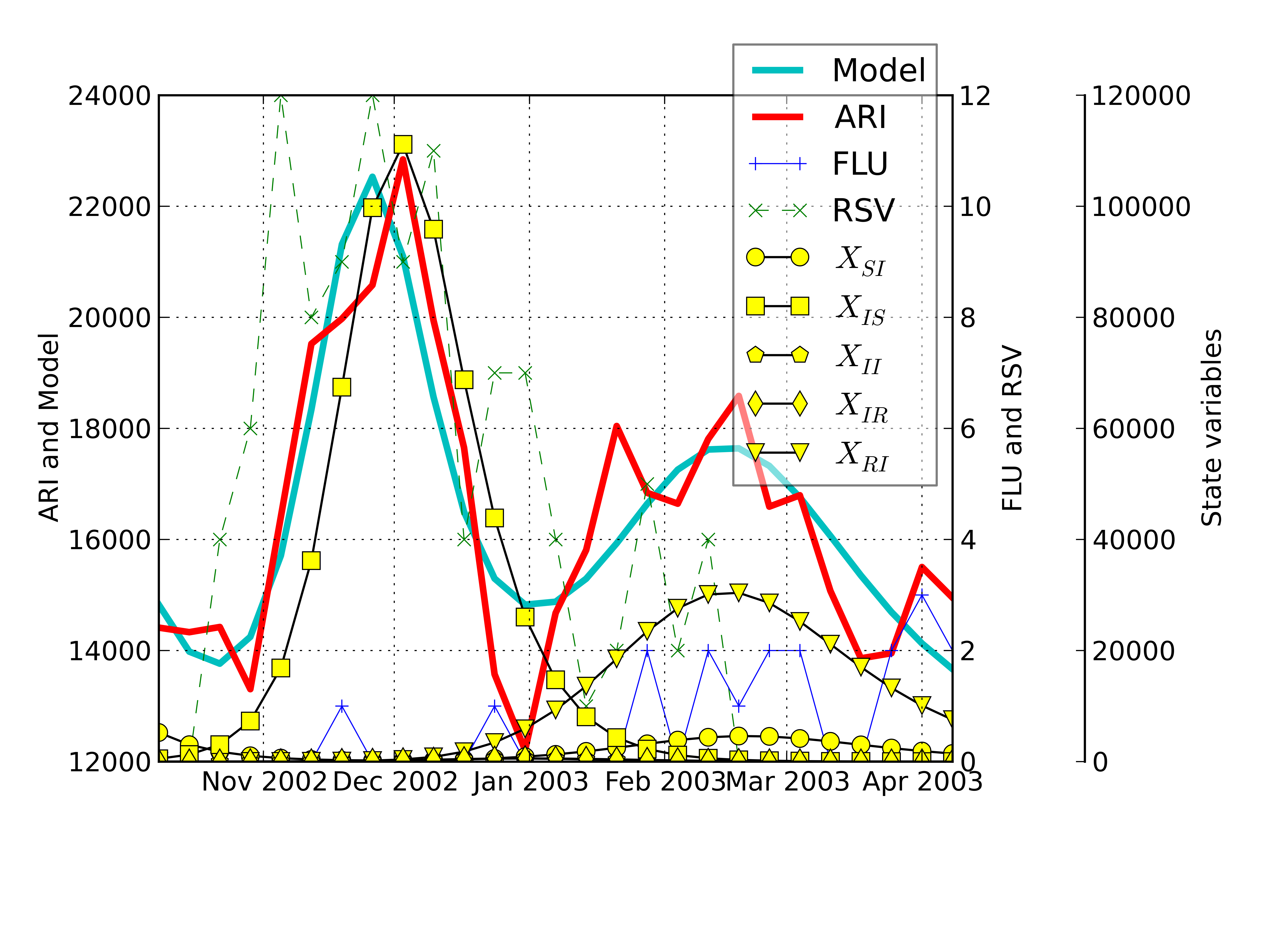}
\caption{{\bf Superinfection.} Surveillance records of the timing of influenza and syncytial respiratory virus circulation are highly correlated with peaks of ARI}
\label{fig:coinfection_2002_2003}
\end{center}
\end{figure}

\begin{figure}[!ht]
\begin{center}
\includegraphics[width=\textwidth]{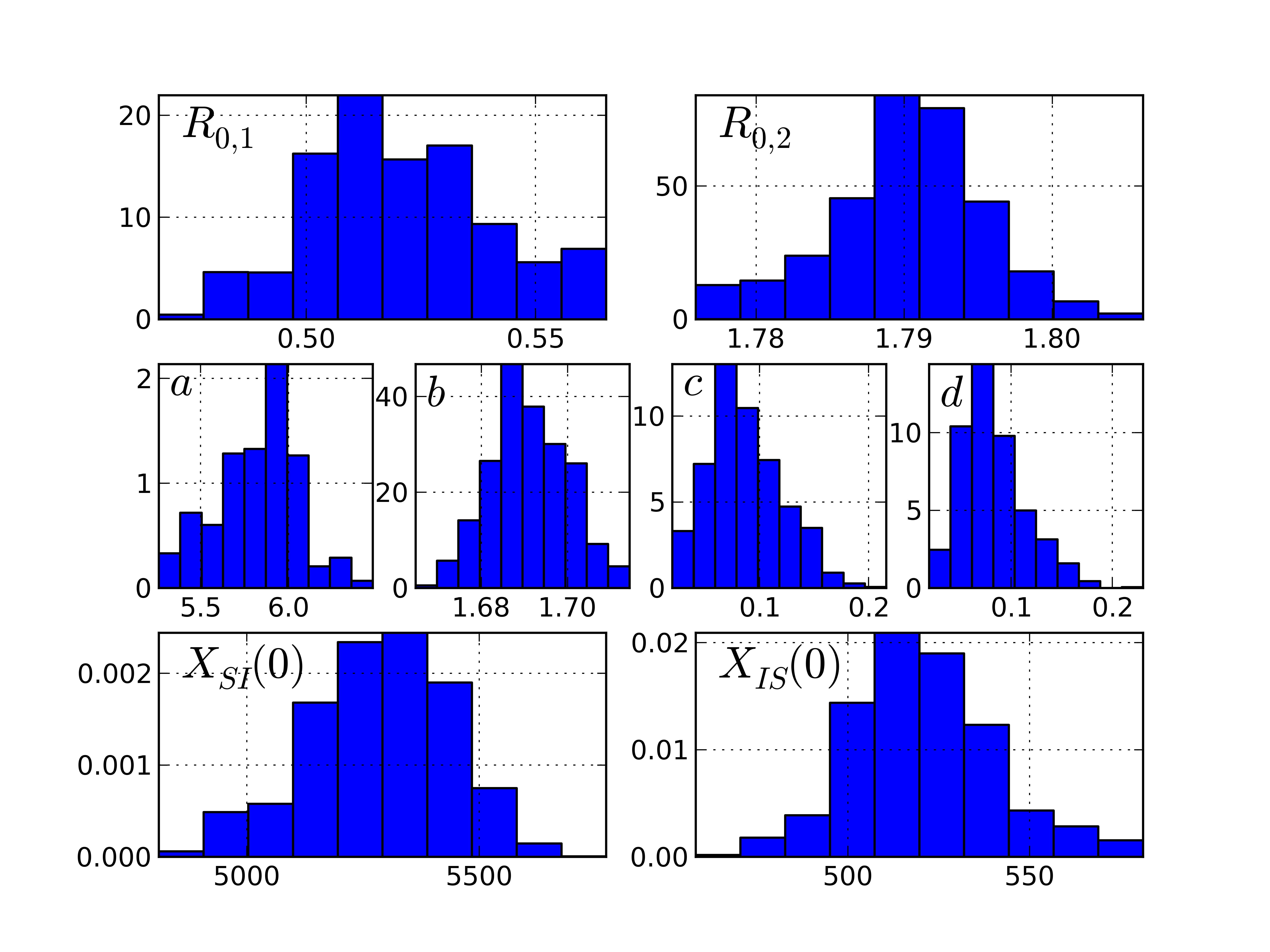}
\caption{{\bf Parameter histograms and superinfection.} The value of the 
estimated parameters support the occurrence of superinfection as the underlying ecological mechanism of interaction between influenza and RSV.}
\label{fig:coinfection_pars_2002_2003}
\end{center}
\end{figure}

Figures~\ref{fig:coinfection_2002_2003} and~\ref{fig:coinfection_pars_2002_2003}
correspond to the 2002-2003 season, where RSV peaked first. 
The comparison of model~(\ref{eq:mean-field}) state variables evaluated at the
maximum a posteriori estimator (MAP) of the parameters,
ARI records and model incidence evaluated at the MAP of the parameters, and 
sentinel records of RSV and influenza are in agreement with the claims made in 
Figure~\ref{fig:replacement_number_2002_2003}. We can interpret these results in 
the following manner, during the first half of the ARI high season RSV has bigger 
fitness than influenza ($R_{0,2}>R_{0,1}$), therefore peaks first. Nevertheless, 
both diseases are present during the ARI high season. 
Figure~\ref{fig:coinfection_pars_2002_2003} shows histograms of the model parameters.
The relative value of the contact rates shows that the model is in the superinfection
regime.

\begin{figure}[!ht]
\begin{center}
\includegraphics[width=\textwidth]{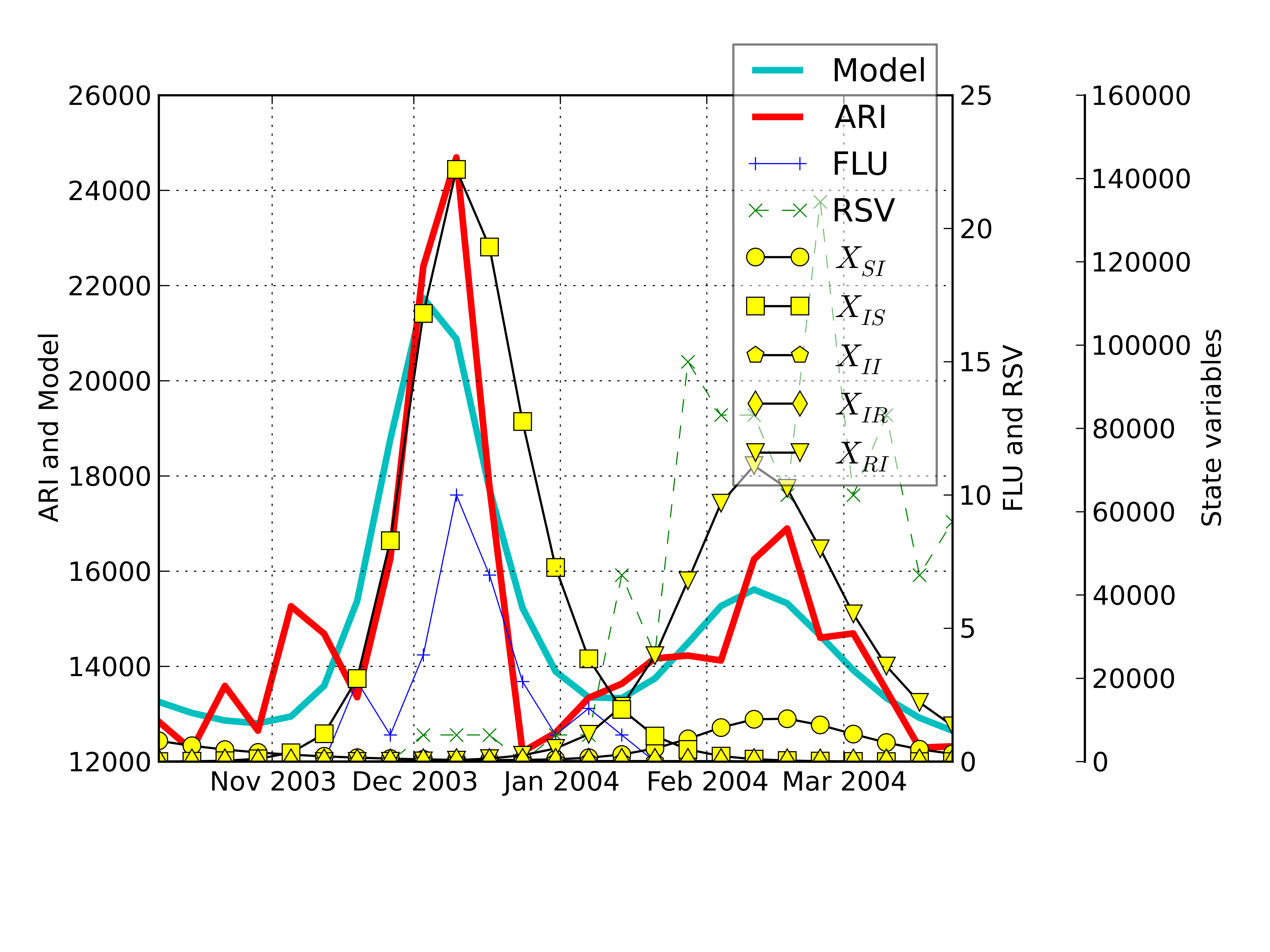}
\caption{{\bf Superinfection.} This season influenza is more fit at the beginning 
of the ARI high season. Further research is necessary in order to determine what
causes switches between influenza and RSV fitness at the beginning of the season.}
\label{fig:coinfection_2003_2004}
\end{center}
\end{figure}

We have obtained similar results to 2002-2003 in all years where a non-pandemic
influenza outbreak occurred, see supplementary material. Of note, there are years 
where RSV peaks first, while the other scenario is also possible (see 
Figure~\ref{fig:coinfection_2003_2004}) 
where influenza peaks first through the same mechanism described above. Remarkably 
in the 2003-2004 season we have posterior marginal distributions
with two modes. The reason is that the epidemic model is symmetric and we haven't incorporated into our inference process the sentinel records of influenza and RSV.
In all cases studied the coinfected state $X_{II}$ remains at very
low level compared with the rest of the infected states. We consider low
counts of $X_{II}$, the delay in the increase in $X_{RI}$ and the larger counts
of $X_{RI}$  compared to $X_{SI}$ as three lines of evidence of viral interference.
We do not have however an explanation for the fact that either, RSV or influenza
have higher fitness at the beginning of the ARI season. A more throughout analysis 
will be reported elsewhere incorporating information related to the influenza 
vaccination coverage and effectiveness and weather covariates.

\section*{Discussion}

Respiratory viruses are common causes of ARI; influenza and RSV epidemics are associated to increasing numbers of consultations and hospitalizations. In temperate climates there is an expected increase in ARI during the winter season. However, the onset, extent and duration of these epidemics are not always predictable. Variability in ARI epidemiology may be associated to environmental factors (such as temperature and pollution), viral factors (diverse etiological agents and strain variations), and host factors (demographic features, population susceptibility). 
In this work we have developed a mathematical model to analyze the interaction between two pathogens (influenza and RSV) as causes of respiratory infections and compared the results of this model to ARI time series obtained during a seven year period in the State of San Luis Potos\'i, as well as virological information regarding influenza and RSV circulation in this region.
The model for each season showed similar results with a bimodal distribution of ARI in which the first peak is explained by infections caused by one pathogen followed by a second outbreak associated to circulation of the second pathogen. The first outbreak could be associated to either influenza or RSV; however, results for every season were similar. Thus, the observed epidemiological features do not appear to depend on the specific agent responsible for the epidemic; rather, it appears that ARI epidemiology depends of the interaction between both agents and the host population resulting in a dominant virus at the start of the peak season which declines during the second part of winter allowing for the second virus to emerge. The model shows that at the onset of the epidemic ARI season both viruses are present in the population; however, during a particular season there are certain characteristics, yet to be defined, that render one virus fitter than the other to establish a disseminated outbreak, resulting in infection of a large proportion of the population. Of note, our model aims to describe the behavior of infections within the whole population (including asymptomatic, mild, and severe infections) and, therefore, a large proportion of the population is expected to be affected by at least one of these two pathogens. In this context, symptomatic infections requiring medical attention comprise only a fraction of all infections occurring during an epidemic period. For instance, almost all children suffer at least one RSV infection during the first two years of life and reinfections are common~\cite{bocchini2009american}. Symptomatic influenza infections severe enough to require outpatient/emergency department medical attention have been estimated to occur in 5.6-12.2\% of children below 5 years of 
age~\cite{poehling2006underrecognized}. In addition, a large proportion of infections are known to be mild (not requiring medical care) or asymptomatic. Ascertainment of asymptomatic infections is more difficult in clinical practice, since serological studies are required. Epidemiological studies have shown that during annual influenza epidemics 25-50\% of the population is infected by this virus~\cite{glezen1996emerging}. The results obtained by this model correlate to the observed ARI time series which are assumed to reflect the seasonality of all infections within the community. In addition, each of the outbreaks described by the model corresponded to either influenza or RSV activity in our region. Of interest, the largest proportion of individuals affected by the second outbreak is predicted to occur in those who have previously been affected (either by symptomatic of asymptomatic infection) by the first virus that circulated in the community; on the other hand the attack rate in those that did not suffer infection by the first pathogen is predicted to be lower. These results would reflect the existence of a group of individuals that are less likely to be infected by both pathogens (either because they are resistant to infection or are less likely to be exposed) while individuals that are affected by a virus may be rendered more susceptible to sequential infection by a second pathogen.
Interestingly, this mathematical model indicates the existence of significant interactions between the two pathogens (either simultaneously of sequentially). 
Of note, our results support the presence of interference between the two viruses 
leading to the occurrence of sequential outbreaks, which in ecological terms is 
called superinfection.  
Of particular interest, the similarities in seasonal trends during each winter, independent of the leading virus at the onset of peak ARI activity may allow to establish a general model that may be applied to subsequent winter seasons 
to aid in outbreak management and may serve to test hypothesis regarding the effectiveness of preventive strategies, such as vaccination.

\section*{Acknowledgments}

We would like to thank the State Health Services of San Luis Potos\'i; particularly 
Dr. Jos\'e de Jes\'us Mendez de Lira who provided us with records of ARI of San Luis 
Potos\'i. We would like to thank Dr. Andreu Comas Garc\'ia for useful discussions,
feedback and insight. Also, the authors would like to acknowledge financial support 
from Fondo Mixto de Fomento a la Investigaci\'on Cient\'ifica y Tecnol\'ogica, 
CONACYT-Gobierno del Estado de Guajanuato, GTO-2011-C04-168776. 

\bibliography{flu_rsv}

\section*{Figures}

\begin{figure}[!ht]
\begin{center}
\includegraphics[scale=0.75]{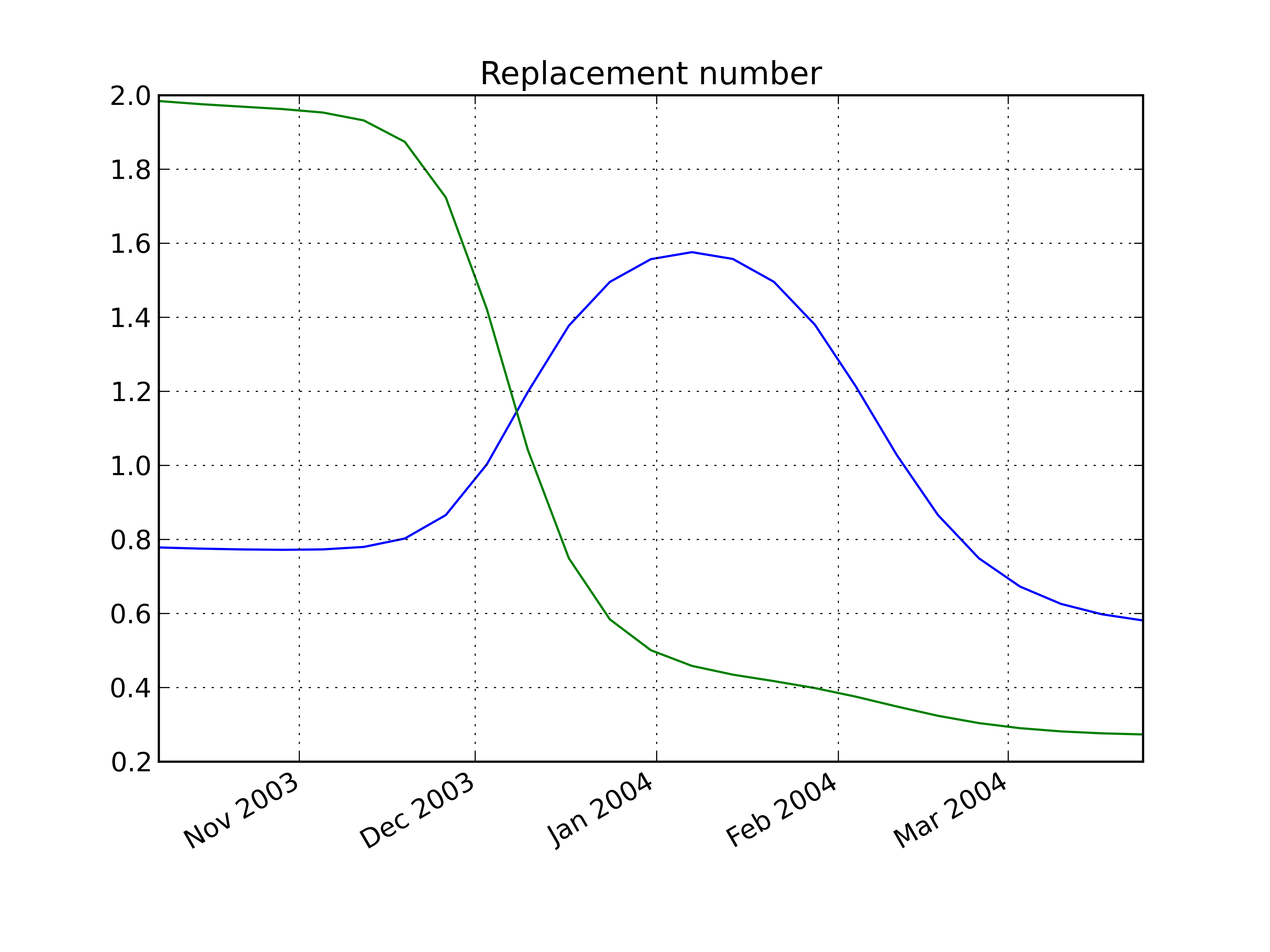}
\caption{{\bf Replacement number as a measure of fitness.} As opposed to 
Figure~\ref{fig:replacement_number_2002_2003}, Influenza 
(green line) has higher fitness at the beginning of the ARI high 
season. Similarly, RSV (blue line) starts with lower fitness but
proliferates once influenza fitness decreases below a certain level (Color online).}
\label{fig:replacement_number_2003_2004}
\end{center}
\end{figure}

%

\end{document}